\documentclass[letterpaper,twocolumn,10pt]{article}
\usepackage{usenix-2020-09}
\pdfoutput=1

\usepackage{tikz}
\usepackage{amsmath}

\usepackage{filecontents}

\usepackage{cite}
\usepackage{amsmath,amssymb,amsfonts}
\usepackage{graphicx}
\usepackage{textcomp}
\usepackage{xcolor}
\usepackage{booktabs}
\usepackage{bm}

\usepackage{algorithm}
\usepackage{algpseudocode}
\usepackage{amsmath}
\usepackage{threeparttable}

\usepackage{authblk}

\begin{document}

\date{}

\title{\Large \bf Directed Greybox Fuzzing with Stepwise Constraint Focusing}


\author[1]{Xiaofan Li}
\author[2]{Xuan Li}
\author[1]{Guangfa Lv}
\author[3]{Yongzheng Zhang \thanks{Corresponding author: zhangyz18177@cetcsc.com}  }
\author[1,4]{Fengyu Wang \thanks{Corresponding author: wangfengyu@sdu.edu.cn}  }
\affil[1]{School of Software, Shandong University, Jinan, China}
\affil[2]{National Computer Network Emergency Response Technical Team/Coordination Center of China, Beijing, China}
\affil[3]{China Electronic Technology Cyber Security Co., Ltd., Beijing, China}
\affil[4]{Quan Cheng Laboratory, Jinan, China}

\renewcommand*{\Affilfont}{\small\it} 
\renewcommand\Authands{ and } 
\date{} 

\maketitle

\begin{abstract}
Dynamic data flow analysis has been widely used to guide greybox fuzzing. However, traditional dynamic data flow analysis tends to go astray in the massive path tracking and requires to process a large volume of data, resulting in low efficiency in reaching the target location. \par

In this paper, we propose a directed greybox fuzzer based on dynamic constraint filtering and focusing (CONFF). First, all path constraints are tracked, and those with high priority are filtered as the next solution targets. Next, focusing on a single path constraint to be satisfied, we obtain its data condition and probe the mapping relationship between it and the input bytes through multi-byte mapping and single-byte mapping. Finally, various mutation strategies are utilized to solve the path constraint currently focused on, and the target location of the program is gradually approached through path selection. The CONFF fuzzer can reach a specific location faster in the target program, thus efficiently triggering the crash. \par

We designed and implemented a prototype of the CONFF fuzzer and evaluated it with the LAVA-1 dataset and some real-world vulnerabilities. The results show that the CONFF fuzzer can reproduce crashes on the LAVA-1 dataset and most of the real-world vulnerabilities. For most vulnerabilities, the CONFF fuzzer reproduced the crashes with significantly reduced time compared to state-of-the-art fuzzers. On average, the CONFF fuzzer was 23.7x faster than the state-of-the-art code coverage-based fuzzer Angora and 27.3x faster than the classical directed greybox fuzzer AFLGo. \par
\end{abstract}

\section{Introduction}

Fuzzing has become one of the most effective techniques for finding vulnerabilities in software programs. Traditional greybox fuzzers usually address the code in a program equally and maximize code coverage to reveal vulnerabilities in different parts of the program. Unlike coverage-guided fuzzing, directed greybox fuzzing (DGF) aims to satisfy the path constraints and reach a given set of program locations precisely, i.e., target locations.\par

DGF is often used in specific cases. For example, when a developer finds a vulnerability in a program and patches the code to fix it, patch testing is required to verify that the fix does not introduce additional vulnerabilities. In this case, DGF can be used to quickly reach the patch location and generate a large number of test cases for patch testing. As another example, crash reports typically contain no user input that triggers the vulnerability due to user privacy reasons. Thus, developers need to reproduce the crash based on the stack track generated by the crash and fix the bugs.\par

The traditional method to reach the target location deep in the program is solving the path constraint with symbolic execution. Essentially, directed symbolic execution (DSE) transforms the reachability problem into an iterative constraint satisfaction problem. In each iteration, DSE identifies branches closer to the target through program analysis and solves constraints using constraint solvers such as Klee\cite{cadar2008klee}, BugRedux\cite{jin2012bugredux}, and QSYM\cite{yun2018qsym}. However, DSE relies on extensive program analysis and complex constraint solving, which greatly reduces its constraint solving efficiency. \par

In recent years, DGF based on control flow information and data flow analysis has shown great potential and attracted much academic attention. DGF turns the reachability problem into an optimization problem. On the one hand, several fuzzers utilize control flow information to enhance the directionality of fuzzing, and AFLGo\cite{bohme2017directed} guides the generation of inputs closest to the target location in the control flow. On the other hand, some fuzzers rely on data flow analysis to locate the constraint in the input. Vuzzer\cite{rawat2017vuzzer} and Angora\cite{chen2018angora} employ taint analysis to identify the input bytes used in the constraint. RedQueen\cite{aschermann2019redqueen} proposes an input-to-state correspondence to solve the magic bytes and checksums in the program by coloring the bytes. Greyone\cite{gan2020greyone} uses data flow sensitive fuzzing-driven taint inference to infer variable taint and monitors the correspondence between input bytes and value changes of the variables. CAFL\cite{lee2021constraint} guides data flow through constraints by generating custom templates of ordered target locations and data conditions. These solutions demonstrate the good performance of utilizing control flow information and dynamic data flow analysis to guide the fuzzing direction and satisfy the data conditions of the constraints. \par

\subsection{Questions to Address}
Traditional data flow analysis requires tracking a large number of irrelevant data flows. On the one hand, dynamic taint analysis needs to track taint execution paths unrelated to the constraints, thus prone to over-taint and under-taint, as in TaintScope\cite{wang2010taintscope} and Angora\cite{chen2018angora}. On the other hand, traditional data flow analysis tracks program variables not tied to the constraints. RedQueen\cite{aschermann2019redqueen} identifies the magic bytes and checksums in the program. Greyone\cite{gan2020greyone} monitors the variables used in program path constraints. However, DGF usually only concerns the path constraints reaching the target location. Tracking and analyzing a large number of data flows irrelevant to the target location reduces the efficiency to reach the target location. Thus, we have the first research question to address. \textbf{RQ1: How to reduce data flow tracking for efficient DGF?} \par

Existing directed greybox fuzzers usually only satisfy specific paths to reach the target locations. For example, AFLGo\cite{bohme2017directed} only calculates the length of the shortest path to the target location as the direction of seed mutation. To speed up crash reproduction, CAFL\cite{lee2021constraint} defines a set of ordered target locations triggering crashes as DGF targets. Other solutions, such as Hawkeye\cite{chen2018hawkeye}, LOLLY\cite{liang2019sequence}, and Berry\cite{liang2020sequence}, propose using the similarity between the execution track of the seed and the target execution track as a measurement for path evaluation. The above approaches have shown good performance in reproducing crashes for DGF. However, in patch testing, covering more paths to the target location is more likely to trigger crashes. Thus, we have the second research question to address. \textbf{RQ2: How to achieve more comprehensive path coverage?} \par

Traditional fuzzers use genetic algorithms to randomize the mutation seeds, such as AFL. Path constraint data conditions of higher complexity often lower the probability that the random mutation satisfies the constraint. Therefore, state-of-the-art fuzzers use data flow analysis to reveal the correspondence between inputs and constraints and assist seed mutation, which include RedQueen\cite{aschermann2019redqueen}, Angora\cite{chen2018angora}, and Greyone\cite{gan2020greyone}. However, precisely inferring the correspondence between inputs and constraints usually requires single-byte mutation, and its cost increases proportionally with the input length. Thus, we have the third research question to address. \textbf{RQ3: How to mutate seeds more quickly and precisely?} \par

\subsection{Our Solution}

We propose a novel DGF solution with constraint filtering and focusing (CONFF) to address the aforementioned questions.\par

\textbf{Constraint filtering and focusing.} During DGF, the complex constraints need to be satisfied step by step to reach the target location, and the full tracking of constraints is a heavy cost. To reduce the cost for data flow analysis, we incorporate constraint filtering into DGF. Constraints are pre-assigned priorities based on their distance to the target location, and those to be satisfied are filtered out as the scheduling target for the next step. Then, we incorporate constraint focusing into data flow tracking. The CONFF fuzzer only focuses on the current path constraint to be satisfied, i.e., the one with the highest priority, while ignoring other constraints. The filtered unique constraint serves as the target constraint and guides the mutation of the input seeds. As only the data flow information generated by the target constraints is required, the amount of information for data flow analysis is greatly reduced. In addition, tracking data flow information through the constraints avoids the labor-intensive efforts of composing taint propagation rules. Only the data flow acquisition rules are required according to the limited number of constraint types. \par

\textbf{Path exploration.} 
To achieve effective directed fuzzing, paths to the target location should be covered comprehensively. Different paths to the target location contain different nodes and, therefore, different constraints. Thus, we focus on the constraints. Based on the function call graph and the control flow graph, the constraints capable of reaching the target location are assigned distance values, i.e., their distances to the target location. With the CONFF fuzzer, constraints are prioritized according to their distance values. While exploring the target location, the constraints are selected and satisfied according to their priorities until finally covering all path constraints. Thus, all paths to reach the target location are tried under the guidance of the constraints.\par

\textbf{Byte Probing.} The CONFF fuzzer can mutate the seeds quickly and accurately according to path constraints, which requires mapping the inputs to the constraints based on the data flow characteristics. Traditional directed greybox fuzzers construct mapping relations through single-byte mutation. Byte-by-byte probing can lead to the waste of time, especially when the input contains a large number of bytes irrelevant to the control flow (e.g., images and videos with large data volumes). Intuitively, data associated with path constraints are usually at specific positions in the inputs. For example, the magic byte marking the file type is usually at the beginning of the input bytes and contiguous. The CONFF fuzzer first obtains the rough mapping of path constraints in the inputs by multi-byte probing and then performs single-byte probing on each byte within the rough mapping so as to quickly and accurately construct the conditional data mapping between input bytes and path constraints. \par

\subsection{Results}
We implemented a prototype of the CONFF fuzzer and evaluated it on the LAVA-1 dataset\cite{dolan2016lava} and 35 real-world crashes.\par

Based on the LAVA-1 dataset, bugs were manually injected into the software. We compared the time required to reproduce the crashes using AFLGo, Angora, and CAFL with that of the proposed fuzzer. Experimental results show that the CONFF fuzzer can quickly reproduce the crashes and even reproduced some of the shallow vulnerabilities injected in less than 1 minute. \par

In terms of the time to reproduce the real-world crashes, the CONFF fuzzer was compared against the classical directed greybox fuzzer AFLGo and the advanced greybox fuzzer Angora based on data flow analysis. The evaluation show that the CONFF fuzzer outperforms AFLGo by 27.3 times and Angora by 23.7 times on average in reproducing common vulnerabilities and exposures (CVEs). Meanwhile, the CONFF fuzzer only spends the same time cost as clang sanitizer to trigger the crashes in some vulnerable program locations. \par

The main contributions of this study can be summarized as follows:
\begin{itemize}
	\item A novel DGF method with CONFF is proposed to improve the efficiency of triggering crashes by addressing only the currently most favorable path constraint to reach the target locations.
	\item We propose a constraint filtering and focusing strategy. The constraints are filtered according to their priorities, which are assigned in advance in the static graphs. The CONFF fuzzer only strives to track and satisfy the constraint with the currently highest priority.
	\item A two-stage mapping method is proposed to detect the relationship between path constraints and input bytes quickly and precisely. First, the rough mapping is obtained by multi-byte probing. Then, the single-byte probing is performed inside the rough mapping to obtain the byte-level mapping.
	\item We implemented a prototype of the CONFF fuzzer and extensively evaluated it on LAVA-1 dataset and real-world open-source applications. The results show that the CONFF fuzzer has significant advantages in directed fuzzing.
\end{itemize}

\section{Design}
\subsection{Overview}
DGF with CONFF aims to satisfy the path constraints and data conditions and reach the target location more quickly. Unlike traditional directed fuzzers that mutate the seeds with distance measurements, the CONFF fuzzer uses distance measurements to filter path constraints. The core process of DGF is as follows. First, it focuses on the path constraint that is capable of reaching and closest to the target location. Second, the data condition of the path constraint is used to guide the seed mutation to satisfy the path constraint. This process is repeated to satisfy the constraints and advance along the path until reaching the target location. \par

\begin{figure*}[htpb]
	\centering
	\includegraphics[width=0.90\textwidth]{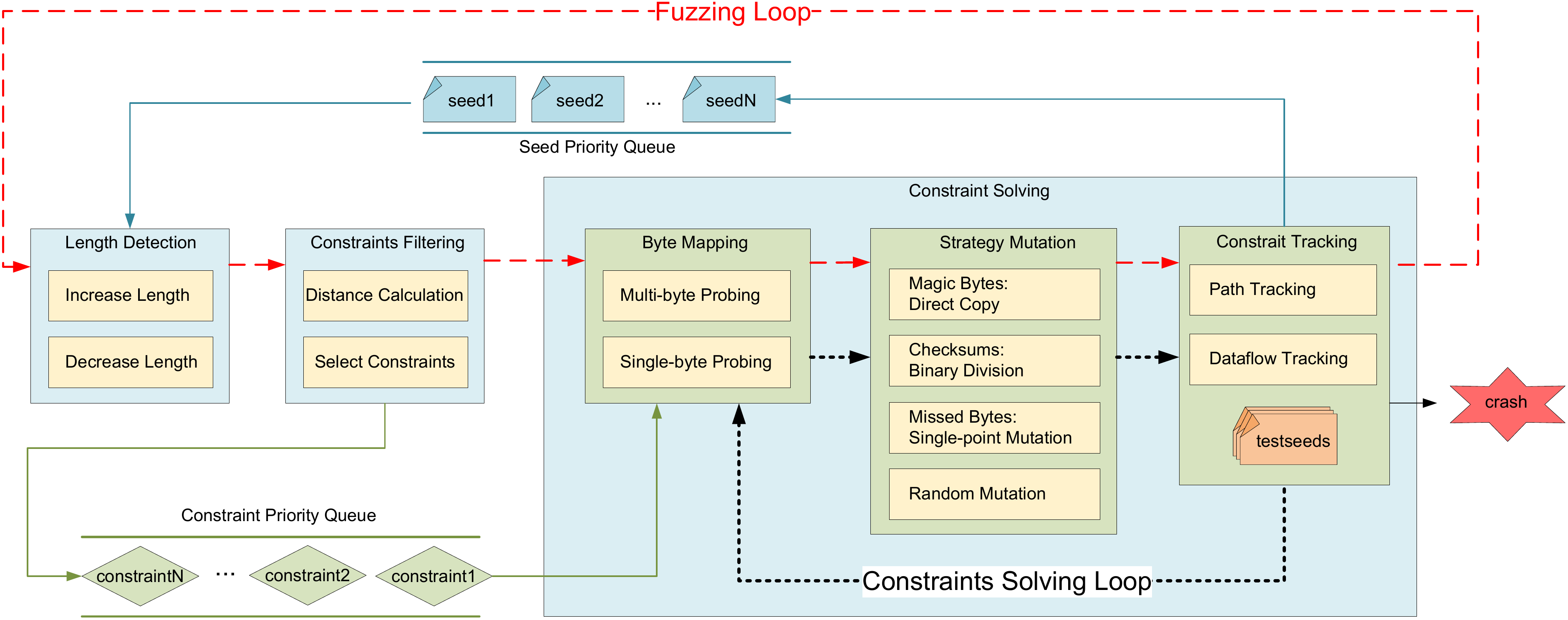}  
	\caption{Architecture of the CONFF fuzzer.}
	\label{architecture}
\end{figure*}

As shown in Figure \ref{architecture}, the fuzzing workflow of the CONFF fuzzer is divided into three stages, including length detection, constraint filtering, and constraint solving. Constraint solving is further divided into three stages, i.e., byte mapping, strategic mutation, and constraint tracking. The CONFF fuzzer maintains two priority queues, the seed queue and the constraint queue. Algorithm \ref{algfuzzingloop} demonstrates the steps in practical operation. \par

The outer fuzzing loop is driven by the seed queue. The seeds in the seed queue are sorted by their priorities, and the one with the highest priority serves as the input. During length detection, the length of the seed is checked and adjusted if necessary. The program is launched with the seed, and the data flow generated by the program is recorded as it traverses the path constraints. In the constraint filtering stage, the distance from the path constraint to be satisfied and the target location is calculated, and the path constraints with the distance measurement are stored in the constraint queue and prepared for constraint solving. \par

The inner constraint solving loop is driven by the constraint queue. The path constraint with the highest priority is selected from the constraint queue to serve as the next solution target. Path constraints with higher priorities are closer to the target location and have greater possibilities of getting closer to the target location after their solution. When solving constraints, the path constraint currently focused on is registered in the tested program, and only its corresponding data flow information is obtained, thus accelerating the data flow analysis by greatly reducing the information. In the byte mapping stage, the mapping relationship between path constraints and seed input bytes are constructed. During the subsequent strategic mutation, different mutation strategies are adopted to mutate the specific bytes in the seed according to the type of path constraints and data conditions. The mutated seed is used as the input of the tested program. The constraint solving results are divided into two cases. If the constraint focused on is solved, the mutated seed is added to the seed queue, and the inner loop is terminated prematurely to start a new branch exploration. If the constraint is not solved, the path constraint is abandoned, and focus shifts to the next path constraint from the constraint queue until the end of the constraint queue. \par

\begin{algorithm}
	\renewcommand{\algorithmicrequire}{\textbf{Input:}}  
	\renewcommand{\algorithmicensure}{\textbf{Output:}} 
	\caption{Two Loop of CONFF Fuzzer}
	\label{algfuzzingloop}
	\begin{algorithmic}[1] 
		\Require
		$program$: test program after instruction;
		$initseeds$: init seeds;
		$graph$: program static graph;
		$location$: sanitizer of crash locations;
		\Ensure
		$report$: execution and crash information
		\Function {MainFuzzer}{$initseeds, graph, location$}
		\State $seed\_queue \gets \{initseeds\}$ // Seed Priority Queue
		\State $cstr\_queue \gets \{\}$ // Constraint Priority Queue
		\State // Fuzzing Loop
		\While{$seed\_queue \neq \varnothing$}
		\State $seed \gets seed\_queue $ \quad // Select highest priority
		\State $seed \gets $ \Call{DetectLength}{$seed$}
		\State $info \gets $ \Call{Executor}{$program, seed$}
		\State $cstr\_queue \gets $ \Call{Distance}{$graph, location, info$}
		\State $flag \gets $ true
		\State // Constraints Solving Loop
		
		\While{$cstr\_queue \neq \varnothing $ \textbf{and} $flag$}
		\State $cstr \gets cstr\_queue$ 
		\State $map \gets $ \Call{Mapping}{$program, seed, cstr$}
		\State $tseed \gets $ \Call{Mutate}{$seed, map$}
		\State $info \gets $ \Call{Executor}{$program, tseed$}
		
		\If{\Call{Satisfy}{$cstr, info$}}
		\State $flag \gets $ false \quad // Early exit
		\State $cstr\_queue \gets \{\}$
		\EndIf
		
		\State $seed\_queue \gets $ \Call{Coverage}{$program, tseed$} 
		\State $report \gets $ \Call{SaveCrash}{$info$}
		\EndWhile
		
		\EndWhile
		
		\EndFunction
	\end{algorithmic}
\end{algorithm}

\subsection{Priority Queue}
\paragraph{Seed Priority Queue}
The seed queue schedules the input seeds sent to the tested program during fuzzing. For directed greybox fuzzers, the quality of seeds is crucial. Thus, we propose a comprehensive measurement of the seed priority. Considering the fuzzing efficiency, the CONFF fuzzer only adds the seed with a higher priority than the current one to the seed queue and discards seeds with lower priorities.

The main goal of directed greybox fuzzers is to reach the target location, where the seed is used as an input to gradually drive the program closer to the target location. The closeness to the target location is the most important indicator to measure seed priority. In this study, we represent this closeness with the seed distance, which can be derived from the path constraints and coincides with the minimum constraint distance in the path constraints reached by the program. A smaller seed distance to the target location means a higher seed priority. \par

Code coverage is an important metric in traditional greybox fuzzing. Different from traditional greybox fuzzing, directed greybox fuzzing pursues to reach the target location quickly. Higher code coverage usually means a higher probability of triggering a crash. In the case of the same seed distance, we choose code coverage as the second metric of seed priority, that is, the higher the coverage, the higher the seed priority. \par

Combining the two metrics mentioned above, we define the seed priority as follows:

\begin{equation}\label{pseed}
	P_{seed}(D,C)=(D*10^8 + \lfloor \frac{1}{C}*10^8 \rfloor) \% 10^{16}
\end{equation}
where $P$ is the seed priority, an integer with a maximum of 16 digits. The first eight digits represent the distance $D$ between the seed and the target location, and the last eight digits represent the inverse of the code coverage $C$. We use $P$ to schedule seeds in the seed priority queue, and a smaller $P$ means a higher priority.

\begin{figure*}[htpb]
	\centering
	\includegraphics[width=0.95\textwidth]{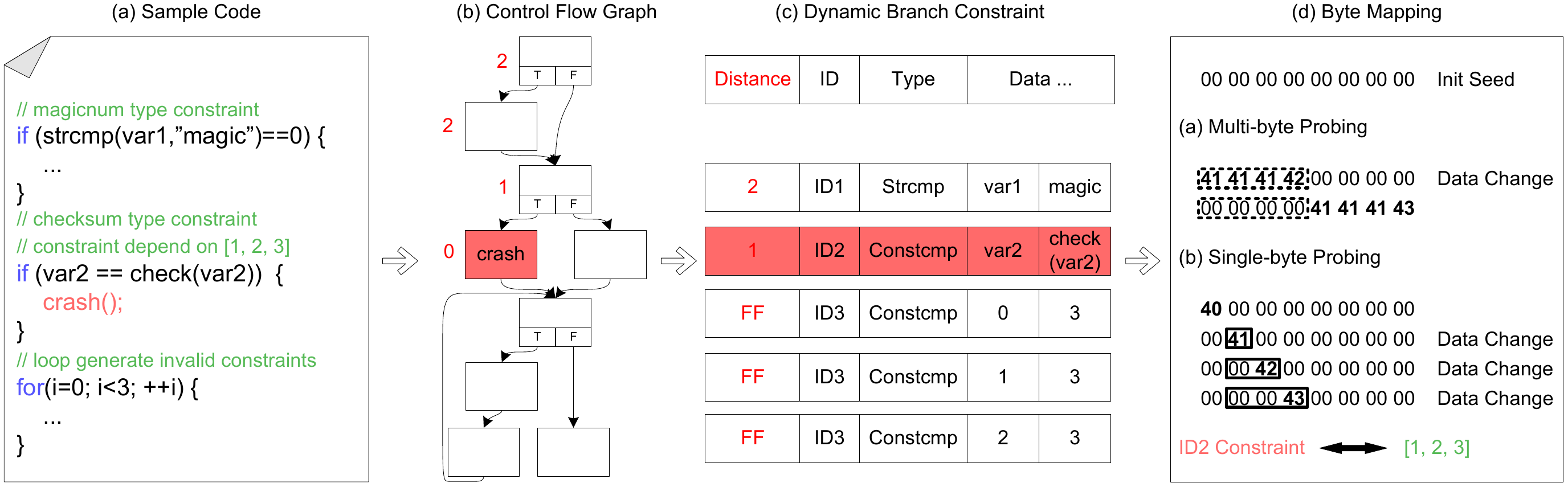}
	\caption{Processing example of the CONFF fuzzer.}
	\label{process}
\end{figure*}

\paragraph{Constraint Priority Queue}
The constraint solving loop maintains a constraint queue storing the valid path constraints filtered from the tested program. Based on the constraint queue, the CONFF fuzzer schedules the constraints to be solved to reach the target location. To determine the next path constraint to be satisfied in the constraint queue, the distance to the target location is usually considered. Therefore, the path constraint priority is assigned based on the constraint distance.

\subsection{Length Detection}
In the length detection stage, constraint coverage or program execution speed can be increased by adjusting the seed length.\par

\paragraph{Increasing Length}
Increasing the seed length can cover more path constraints. On the one hand, some path constraints are the input length, which are only satisfied by the increased seed length. For example, when validating phone numbers, the length of the input phone number needs to be verified. On the other hand, some path constraints are seed length requirements to track their data flow. 
Assuming that the seed length is 0, the path constraints of reaching the target location cannot be covered by feeding the empty seed into the tested program. Hence, it is necessary to increase the seed length and explore the program further. \par

Note that the seed length cannot be increased indefinitely. In general, increasing the seed length is senseless if the path constraint coverage for the tested program remains unchanged. However, some programs (e.g., the base64 program) have looped inputs such that the covered path constraints increases with seed length. Therefore, it is necessary to limit the range of seed length increases. In each new fuzzing loop, the seed length can be increased to twice that in the previous loop through random replication. \par

\paragraph{Reducing Length}
Reducing the seed length can accelerate the execution of the tested program. If increasing the seed length does not track more path constraints, then the seed length is excessively long. In this case, the seed length can be reduced without affecting the number of constraints covered. In this way, irrelevant input bytes can be trimmed to accelerate the tested program in subsequent fuzzing. \par

\subsection{Constraint Filtering}
In the constraint filtering stage, the constraint to be satisfied in the next step is selected by collecting data flow information of the path constraints. \par

\paragraph{Distance Calculation}
The distance to the target location is the basis for directed greybox fuzzers, and accurately obtaining each constraint distance is critical for reaching the target location. \par

We first perform distance measurements on static code blocks. Generally, static code blocks capable of reaching the target location in the static graph are more likely to reach the target location during dynamic execution. To obtain the distance from the basic block to the target location, we obtain the function call graph and control flow graph based on the debug information added during compiling. After obtaining the target location, we employ the breadth-first-search (BFS) algorithm to traverse all the parent nodes on the static graph capable of reaching the target location and attach distance values to the code blocks starting from 0 in the traversal order. In this way, all the basic blocks in the static graph capable of reaching the target location are measured using the distance values. \par

Path constraints are distributed on the code blocks, and the distance of the code blocks for static analysis is the same as that of the path constraints for dynamic execution. 

During directed fuzzing, the constraints are selected and satisfied according to their priorities so as to cover as many constraints capable of reaching the target location as possible. Meanwhile, different path constraints produce different paths, and covering more constraints means trying as many paths as possible. As shown in Figure \ref{process}(b), crashes are selected as the target location after constructing the corresponding control flow graph with the sample code. All parent nodes are traversed via the BFS algorithm starting from the target location, and the distance values are assigned. \par

We abandon the harmonic mean over multiple targets that could lead to inaccurate distance measurements. Instead, we divide the multiple targets into multiple individual targets and perform extensive fuzzing on each of them. Thus, the distance measurements are more accurate, but the time cost of fuzzing is increased. We record the data conditions under which the fuzzing satisfies the constraints. Thus, the prior experience is reused when fuzzing multiple targets, and the time cost for separating multiple targets is compensated. \par

The possible implicit control flow in the tested program could lead to dynamic execution paths inconsistent with the static paths. The number of basic code blocks in the static graph is determined. The dynamic execution path induced by the implicit control flow always lands on a basic code block with a distance value. The implicit control flow problem can be solved by only treating it as a new path that can reach the target location. We need not spend additional cost to track the dynamic paths generated by the implicit control flow.\par

\paragraph{Constraint Filtering}
As a large number of path constraints can be obtained during dynamic tracing, we improve the efficiency by selecting the first three path constraints with the smallest distances as candidates for the next path constraint to be solved. For example, if the obtained constraints have distances of 1, 3, 5, 7, and 9, we select the first three with distances of 1, 3, and 5 as the candidates. The invalid path constraints are marked and filtered out. In general, the constraints generated by the loop statement comparison are invalid. Meanwhile, the path constraints tracked from the function calls in the code block have the same distance as the code block calling the function.  As shown in Figure \ref{process}(c), five corresponding path constraints are tracked for the sample code, among which ID1 and ID2 are generated by conditional control statements. At the same time, ID3 is an invalid path constraint generated by a loop statement. The path constraints with distances are sent to the constraint priority queue, and the closest one to the target location is selected as the next path constraint to focus on and solve.

\subsection{Byte Mapping}
\begin{figure}[htpb]
	\centering
	\includegraphics[width=0.45\textwidth]{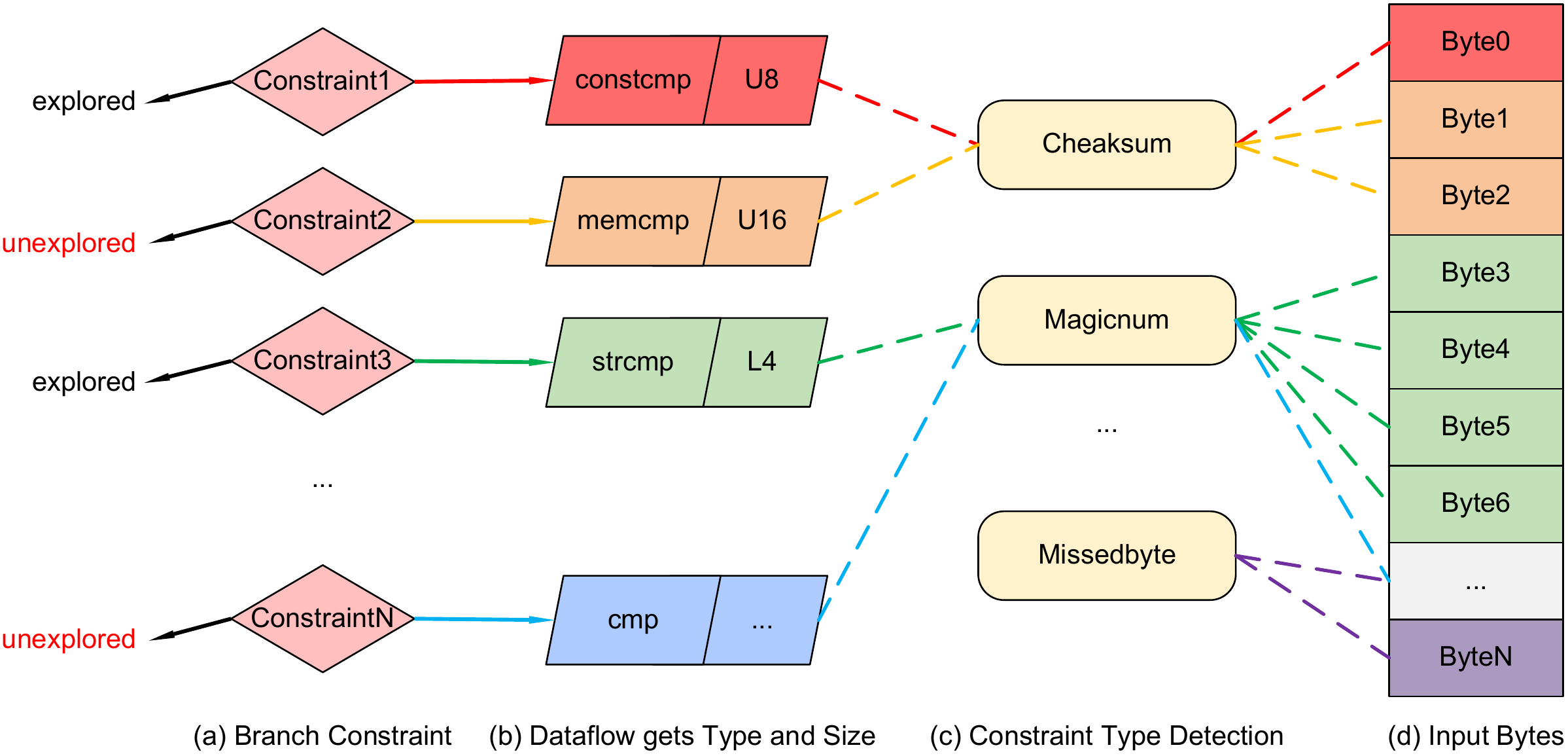}  
	\caption{Mapping between path constraints and seed bytes.}
	\label{compare_map}
\end{figure}

After determining the constraint to focus on and solve, seed mutation is required to satisfy the corresponding data conditions. Instead of random seed mutations, the CONFF fuzzer mutates the exact bytes of the seed corresponding to the path constraint's data conditions. For this purpose, we need to pre-acquire the mapping relationship between the path constraint's data conditions and seed bytes. 
Take the \textit{strcmp(str1, str2)} as an example, where \textit{strcmp} is the path constraint, and \textit{str1} and \textit{str2} are the data conditions, selecting the input bytes alters \textit{str1} and \textit{str2}.
Traditional single-byte mapping is time-consuming. To accelerate the search for the associated bytes, we employ multi-byte mapping in addition to traditional single-byte mapping and propose a two-level byte mapping method. The mapping between path constraint's data conditions and seed bytes is established based on the type of path constraints and the length of path constraint's data conditions (e.g., the lengths of \textit{str1} and \textit{str2} in \textit{strcmp(str1, str2)}, which are obtained from data flow analysis, as shown in Figure \ref{compare_map}.

\textbf{Multi-byte Probing.}
Multi-byte probing reveals the potential relevant seed bytes between mappings. The data conditions corresponding to the path constraints are usually within consecutive seed bytes. For example, executable files' magic bytes (7F 45 4C) identifying their types are usually consecutively distributed in the file header. We split the seed into $m$ segments, each containing the same number of bytes, and mutate all bytes of each segment to check whether the data conditions change in response to byte changes.
The irrelevant byte changes are divided into two cases. In the first, the data condition of the path constraint does not change. In the second, the byte changes render the constraint untrackable. The reason is that such a constraint may be the internal constraint of another constraint of the program, and the mutated byte makes the internal constraint untrackable if the outer constraint is broken.
In contrast, if one data condition changes, the corresponding byte-segment is recorded as potentially related to that data condition. The loop is scaled down to divide the potentially related byte-segment into $m$ parts again until the byte-segment length is below the preset value. At this point, the correlation between the data condition and the seed byte-segment is established. \par

As shown in Figure \ref{process}(d), mutating the first four bytes of the initial seed alters the data condition, but mutating the last four bytes does not. Then, the first four bytes of the seed marked with dotted lines in the figure are considered potentially relevant bytes to the path constraint currently focused on.

\textbf{Single-byte Probing.}
Single-byte probing reveals the exact byte in the mapping. It requires further mutation of each potentially related byte identified by multi-byte probing to precisely determine the mapping relationship between data conditions and seed bytes. If the path constraint's data condition changes along with the byte mutation at a single location, this byte is considered to affect the data condition and recorded as a direct mapping byte to the data condition. If the data condition remain unchanged or the path constraint is not tracked, this byte is deemed irrelevant to the path constraint and ignored as invalid bytes. Thus, the precise mapping between data conditions and seed bytes can be obtained. \par

As shown in Figure \ref{compare_map}, single-byte probing is applied to the potentially related bytes, and the solid line records the exact positions of the mapped bytes. Finally, it can be concluded that the data condition of the path constraint ID2 depends on the 1st, 2nd, and 3rd bytes of the seed. \par

\subsection{Strategic Mutation}
In the probing stage, the mapping relationship between the path constraint's data condition and seed input bytes is constructed. Strategic mutation is applied to quickly satisfy the path constraint's data conditions. After determining the path constraint to be satisfied, the type of path constraint and the length of the data condition are obtained through data flow analysis. Usually, path constraints are divided into binary-path constraints (e.g., \textit{memcmp}) and multiple-path constraints (e.g., \textit{switch}). \par

For different path constraints, the conditions to be satisfied and the branches to be triggered are different. Each condition in the multiple-path constraint represents a new path, and we split the multiple-path constraint into multiple binary-path constraints. 
Then, the different binary path constraints are satisfied to try the different paths. Next, the type of data condition was determined according to the change of the byte in the data condition.\par

We divided the data conditions into three types, namely, fixed-to-fixed, fixed-to-mapped, and mapped-to-mapped:
\begin{itemize}
	\item[$\bullet$] For the fixed-to-fixed type, the path constraints do not change with the seed, thus generally considered invalid and ignored during the selection.
	\item[$\bullet$] For the fixed-to-mapped type, which is usually called magic bytes, the value of the magic byte is directly copied to the seed to quickly satisfy path constraints.
	\item[$\bullet$] For the mapped-to-mapped type, which is usually called checksums, the bytes at both ends of the data condition are treated as a whole, and a distance-difference-guided binary enumeration is used to satisfy the constraint.
\end{itemize}
If none of the above methods can satisfy the constraints, a random mutation strategy is tried on the data conditions. \par

\textbf{Direct copying for magic bytes.}
The path constraints of magic bytes feature one data condition of fixed value and the other changing with the change of seed byte. Such path constraints typically appear in numeric comparisons and string equality comparisons. For numeric-type magic bytes, the value is converted differently depending on the input type, such as converting number to hexadecimal characters or bitwise splitting number into characters. The converted characters are copied directly to the mapped bytes in the seed to satisfy the path constraint's data conditions. For character-type magic bytes, the value of the string is copied directly to the mapped bytes of the seed so that the data conditions required by the path constraint can be satisfied. Note that two possibilities of big-endian and little-endian should be tried.

\textbf{Binary enumeration for checksums.}
The path constraints of checksums feature two data conditions changing simultaneously with the seed input bytes.
A path constraint is satisfied if its two data conditions are equal. 
The difference between the two data conditions can be represented by a binary number.
Assume that the constraint has $n$ binary digits, we generate a numerical sequence of $\{2^{n-1}, 2^{n-2},...,2^1,2^0\}$ as the binary enumeration.
When mutating the corresponding seed position to satisfy a constraint, each value in the binary enumeration sequence is added or subtracted the value from the seed position, and the difference between the two data conditions is used to guide the seed mutation. Specifically, if the difference becomes smaller, the operation result is kept, otherwise the operation result is discarded. This operation is iterated for all values in the binary enumeration sequence until the constraint is satisfied.

\textbf{Single point mutation for missed bytes.}
The path constraint fuzzing covers most of the input bytes. However, some of the bytes in the seed are associated with the implicit control flow and not covered by the path constraint.
These missed bytes may affect the implicit control flow and render the target location unreachable.
Thus, we mutate each missed byte and save the seeds with higher priorities after mutation or the ones triggering new paths.

\textbf{Random mutation.}
If the distance to the target location has not decreased for a long time, the CONFF fuzzer no longer measures seed priority with the distance to the target location. Instead, the seeds are randomly mutated so that the seed covers new code blocks, and the current state is altered.

\subsection{Constraint Tracking}
The mutated seeds are sent to the tested program for execution.
The CONFF fuzzer tracks the dynamic execution path of the tested program and the data conditions of the constraints focused on, and data are collected through shared memory. After constructing the precise mapping between the data condition of the path constraint and the seed byte, the data condition can change along with the seed byte. The CONFF fuzzer measures the difference between data conditions and mutates the seed to reduce it. If the distance to the target location is decreased following the execution path, the path constraint focused on is satisfied. Then, the corresponding mutated seed is added to the seed queue, the constraint solving loop is terminated, and the next loop of fuzzing is initiated. \par

\begin{table}[h] 
	\centering 
	\fontsize{7}{10}\selectfont    
	\caption{Crash reproduction on the LAVA-1 dataset.} 
	\label{LAVA-1} 
	\begin{tabular}{cccccc}
		\toprule
		ID & Crash Location & AFLGo & Angora & CAFL & CONFF \\
		\midrule
		292 & cdf.c:316 & \textgreater24h & \textbf{11s} & 36s & 13s \\
		357 & cdf.c:298 & \textgreater24h & 1h12m55s & 1m18s & \textbf{7s} \\
		660 & softmagic.c:1208 & \textgreater24h & 43s & 36s & \textbf{9s} \\
		1199 & softmagic.c:1762 & \textgreater24h & \textgreater24h & \textbf{18s} & 24s \\
		2285 & funcs.c:65 & \textgreater24h & 42m13s & 42s & \textbf{21s} \\
		2543 & softmagic.c:715 & \textgreater24h & 1h59m48s & 18s & \textbf{5s} \\
		3089 & softmagic.c:1097 & \textgreater24h & 19m2s & 48s & \textbf{8s} \\
		3377 & funcs.c:66 & \textgreater24h & 9m45s & 42s & \textbf{9s} \\
		4049 & funcs.c:62 & \textgreater24h & 17m32s & \textbf{36s} & 1m1s \\
		4383 & softmagic.c:692 & \textgreater24h & \textgreater24h & 12s & \textbf{6s} \\
		4961 & apprentice.c:3222 & \textgreater24h & 10h42m1s & 12s & \textbf{5s} \\
		7002 & softmagic.c:879 & 4m16s & 52s & \textbf{6s} & 15s \\
		7700 & funcs.c:63 & \textgreater24h & 2m9s & \textbf{18s} & \textbf{18s} \\
		9763 & funcs.c:63 & \textgreater24h & \textgreater24h & \textbf{18s} & 27s \\
		13796 & readelf.c:500 & \textgreater24h & \textgreater24h & 121m30s & \textbf{7m29s} \\
		14324 & magic.c:310 & \textgreater24h & 1m33s & 54s & \textbf{6s} \\
		16689 & apprentice.c:493 & \textgreater24h & 18m37s & 42s & \textbf{7s} \\
		\bottomrule
	\end{tabular}
\end{table}

\begin{table*}[h] 
	\centering 
	\fontsize{8}{10}\selectfont    
	\caption{Reproduction of real-world crashes.} 
	\label{CVE} 
	\begin{tabular}{llllcccrr}
		\toprule
		Source & Program & Crash Location & Crash Type & AFLGo & Angora & CONFF & \multicolumn{2}{c}{Speedup} \\
		\midrule
		CVE-2015-5221 & jasper & jas\_tvp.c:111 & use after free & 46s & 1m29s & \textbf{14s} & 3.3x & 6.4x \\
		CVE-2016-4487 & c++filt 2.26 & cplus-dem.c:1239 & use after free & \textbf{2m} & *  & 3m33s & 0.6x & * \\
		CVE-2016-4488 & c++filt 2.26 & cplus-dem.c:1244 & use after free & 11m & * & \textbf{4m28s} & 2.5x & * \\
		CVE-2017-10686 & nasm 2.14rc0 & preproc.c:1305 & use after free & * & * & \textbf{37m36s} & * & * \\
		
		CVE-2019-6455 & rec2csv 1.8 & rec-comment.c:49 & double free & 7s & \textbf{5s} & 1m & 0.1x & 0.1x \\
		
		CVE-2016-4490 & c++filt 2.26 & cp-demangle.c:1596 & buffer overflow & 1m33s & * & \textbf{6s} & 15.5x & * \\
		CVE-2016-4492 & c++filt 2.26 & cplus-dem.c:3781 & buffer overflow & \textbf{9m} & * & 10m31s & 0.9x & * \\
		CVE-2016-4493 & c++filt 2.26 & cplus-dem.c:3789 & buffer overflow & 33m22s & * & \textbf{12m59s} & 2.6x & * \\
		
		CVE-2017-7210 & objdump 2.26 & libbfd.c:184 & heap buffer overflow & \textgreater24h & \textgreater24h & \textbf{4m37s} & 311.9x & 311.9x \\ 
		CVE-2017-11732 & swftophp 0.4.8 & decompile.c:104 & heap buffer overflow & \textgreater24h & 53m52s & \textbf{43m24s} & 33.2x & 1.2x \\
		CVE-2017-11734 & swftophp 0.4.8 & decompile.c:2864 & heap buffer overread & \textgreater24h & \textgreater24h & \textbf{43m38s} & 33.0x & 33.0x \\ 
		CVE-2021-4214 & listswf 0.4.8 & read.c:201 & heap overflow & 4h38m3s & 53m36s & \textbf{22m41s} & 12.3x & 2.4x \\
		
		CVE-2018-8807 & swftophp 0.4.8 & decompile.c:237 & stack overflow & \textbf{2h59m28s} & \textgreater24h & 7h38m19s & 0.4x & 3.1x \\
		CVE-2016-4489 & c++flit 2.26 & cplus-dem.c:4839 & integer overflow & 3m & * & \textbf{39s} & 4.6x & * \\ 
		
		CVE-2017-16883 & swftophp 0.4.8 & outputscript.c:1429 & NULL pointer dereference & 2m59s & \textbf{6s} & 1h42m34s & 0.0x & 0.0x \\
		CVE-2017-8392 & objdump 2.28 & section.c:1395 & NULL pointer dereference & \textgreater24h & \textgreater24h & \textbf{34m55s} & 41.2x & 41.2x \\
		
		CVE-2017-11731 & swftophp 0.4.8 & decompile.c:868 & invalid memory read & 1h16m3s & 53m59s & \textbf{27m22s} & 2.8x & 2.0x \\
		
		CVE-2019-7582 & listswf 0.4.8 & read.c:252 & memory allocation failure & \textgreater24h & \textgreater24h & \textbf{55m29s} & 26.0x & 26.0x \\
		\midrule
		\textbf{Average} &  &  &  &  & \multicolumn{4}{r}{AFLGo:27.3x / Angora:23.7x} \\
		\bottomrule
	\end{tabular}
	\begin{tablenotes}
            \footnotesize
            \item[*] * The program under test failed to launch. 
    \end{tablenotes}
\end{table*}

\begin{figure}[htpb]
	\centering
	\includegraphics[width=0.45\textwidth]{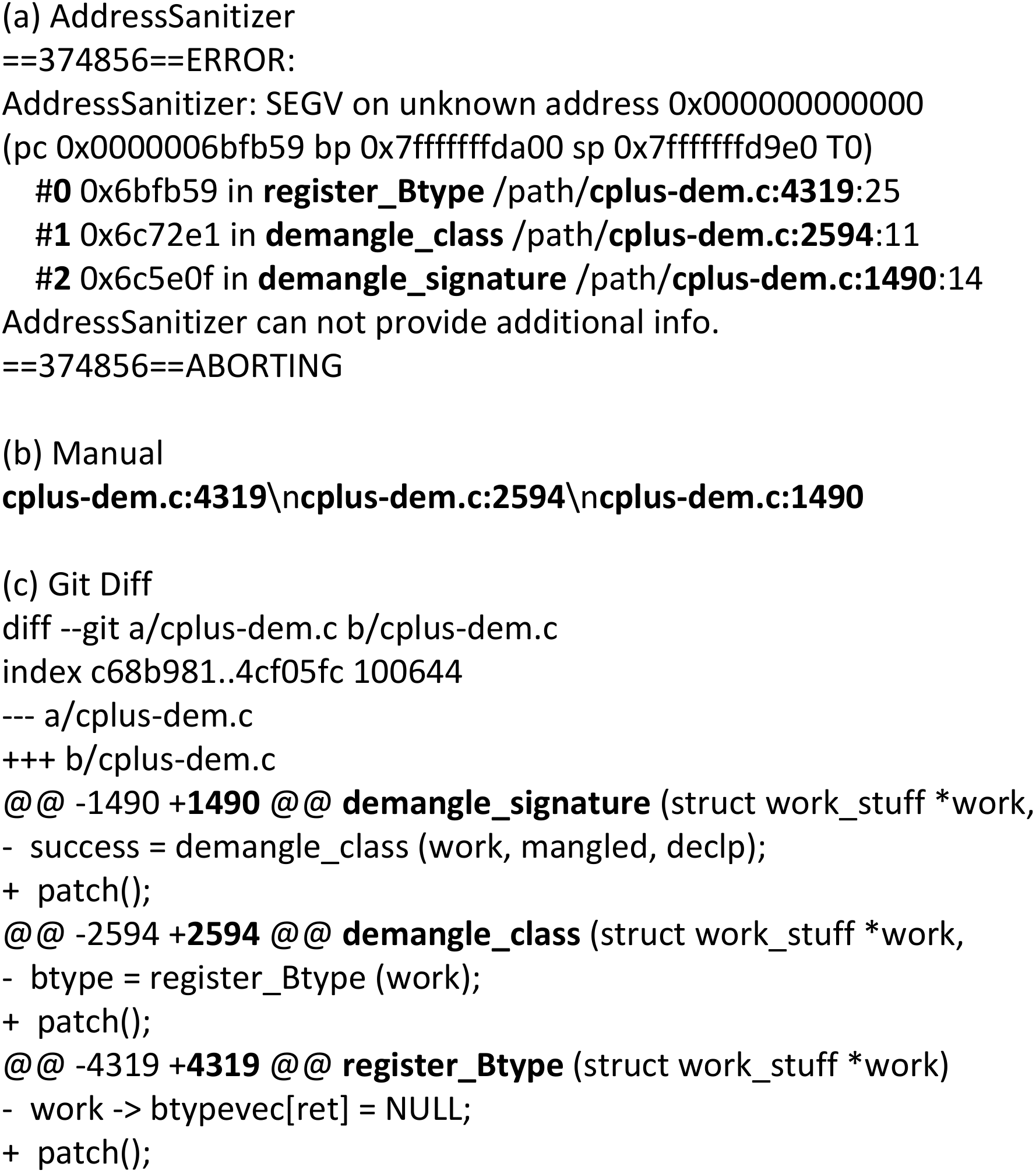}  
	\caption{The sources of crash location information.}
	\label{patchloc}
\end{figure}

\section{Implementation}
We implemented a prototype of the CONFF fuzzer to verify its feasibility. The instrumentation code is written in C, and the core code is written in Python. Meanwhile, the Clang Sanitizer is employed to compile the program while collecting the detailed information about crashes during program execution. The detailed implementation of the system is as follows. 

\textbf{Program Compiling.} We use LLVM wrappers, such as \textit{wllvm}\cite{wllvm} and \textit{gllvm}\cite{gllvm}, to build the LLVM IR bitcode of the tested program. During compiling, the debug information of the program (-g) is extracted, and all optimization options (-O0) are turned off to prevent the optimization of the target location. Subsequently, the LLVM IR bitcode is compiled by using the compiler frontend Clang to add \textit{sanitizer} and \textit{block-level code coverage}. Then, the hook functions of the path constraints are compiled into a static link library and linked into the final program to be tested.

\textbf{Construction of Control Flow Information.}
We use LLVM to generate program call graphs (CGs) and control flow graphs (CFGs) and mark target locations in static graphs based on the program debugging information. Then, the \textit{networkx}\cite{networkx} library is employed to generate the static graphs with distance information. During distance calculation, we only select one target position at a time and use the BFS algorithm to traverse all parent nodes starting from the target location to mark the distance. The graphs with distance information are recorded for fuzzing.

\textbf{Tracking of Data Flow Information.} On the one hand, we track the execution path of the tested program and use the \textit{SanitizerCoverage}\cite{SanitizerCoverage} provided by Clang to build the association between static graph and program execution. The execution path is obtained when the seed is input into the tested program. On the other hand, we track the path constraint's data conditions, and only path constraints generated by constraint focusing are tracked. For efficiency, we use the shared memory to pass information about the path constraint's data conditions. \par

The shared memory is divided into four parts: \textit{path constraint ID}, \textit{code coverage number}, \textit{code execution path}, and \textit{data flow information}. The \textit{path constraint ID} stores the unique identifier of the path constraint to be tracked. The tracking function focuses on the path constraint according to this ID. The unique function address generated during the dynamic execution of the path constraint is used as the \textit{path constraint ID} for distinction. We use the hook tracing function to trace the path constraint. At this point, two function addresses are generated: the call address generated by the function calling the path constraint and the return address generated by the returned result of the path constraint itself. With Address Space Layout Randomization turned off, the \textit{path constraint ID} is uniquely determined by its return address and call address. \par

The \textit{data flow information} is tracked and saved according to the \textit{path constraint ID} to guide the seed mutation for the directed greybox fuzzer. Only the data condition uniquely corresponding to the \textit{path constraint ID} is tracked and saved, while other data conditions in the tested program are ignored, thus greatly reducing the data flow analysis workload and memory consumption. \par

\textbf{Crash Type Distinction.}
We use Clang Sanitizer as compiler for the program, which can detect the crash type and return the call stack at the moment of program crash. Thus, we distinct and record the crashes based on their type and the call stack.

\begin{table*}[h] 
	\centering 
	\fontsize{9}{10}\selectfont    
	\caption{Reproduction of discovered crashes.} 
	\label{newCrash} 
	\begin{tabular}{llllccc}
		\toprule
		Source & Program & Crash Location & Crash Type & AFLGo & Angora & CONFF \\
		\midrule
		mjs-issues-73 & mjs-bin & mjs.c:13679 & use after free & \textgreater24h & * & \textbf{11m51s} \\
		giflib-bugs-74 & gifsponge 5.1 & egif\_lib.c:771 & double free & \textgreater24h & 9m54s & \textbf{3s} \\
		github issues\#149 & swftophp 0.4.8 & read.c:227 & integer overflow & 3m17s & \textbf{33s} & 2m40s \\
		github issues\#76 & swftophp 0.4.8 & decompile.c:2369 & heap buffer overflow & \textbf{20m11s} & \textgreater24h & 7h24m10s \\
		github issues\#79 & swftophp 0.4.8 & outputscript.c:1440 & heap buffer overflow & 2h27m10s & \textgreater24h & \textbf{1h13m57s} \\
		github issues\#78 & swftophp 0.4.8 & decompile.c:629 & NULL pointer dereference & \textbf{9m41s} & 53m56s & 29m13s \\
		github issues\#84 & swftophp 0.4.8 & decompile.c:1193 & invalid memory read & \textgreater24h & \textgreater24h & \textbf{1h20m14s} \\
		github issues\#349 & nasm 2.14rc0 & preproc.c:3868 & invalid memory read & * & * & \textbf{31s} \\
		github issues\#85 & listswf 0.4.8 & blocktypes.c:145 & memory allocation failure & \textgreater24h & \textgreater24h & \textbf{22m41s} \\
		
		- & swftophp 0.4.8 & decompile.c:654 & heap buffer overflow & \textgreater24h & \textgreater24h & \textbf{1h29m43s} \\
		- & mjs-bin & mjs.c:13671 & heap buffer overflow & \textgreater24h & * & \textbf{1m41s} \\
		- & nasm 2.14rc0 & preproc.c:1227 & heap buffer overflow & * & * & \textbf{21m48s} \\
		- & tiffcp 0.4.7 & tiffcp.c:784 & heap buffer overflow & \textgreater24h & \textgreater24h & \textbf{57m34s} \\
		- & swftophp 0.4.8 & main.c:111 & allocation size overflow & \textgreater24h & \textgreater24h & \textbf{6s} \\
		- & mjs-bin & mjs.c:9320 & invalid memory read & \textgreater24h & * & \textbf{13s} \\
		- & swftophp 0.4.8 & decompile.c:1238 & invalid memory read & \textbf{2h39m36s} & \textgreater24h & 6h52m45s \\
		- & c++filt 2.26 & cplus-dem.c:2744 & undefined behavior & \textgreater24h & * & \textbf{21s} \\
		\bottomrule
	\end{tabular}
	\begin{tablenotes}
            \footnotesize
            \item[*] * The program under test failed to launch. 
    \end{tablenotes}
\end{table*}

\section{Evaluation}
In this section, we evaluate the effectiveness of the CONFF fuzzer in reproducing targeted crashes and compare it with other fuzzers. Some performance indexes of the CONFF fuzzer are presented.

\subsection{Experiment Setup}\label{ES}
\textbf{Baseline fuzzers.} We compared the CONFF fuzzer with several closely related fuzzers, including AFLGo\cite{bohme2017directed}, Angora\cite{chen2018angora}, and CAFL\cite{lee2021constraint}. As a classic AFL\cite{AFL}-based directed fuzzer, AFLGo measures the distance to the target location and tries to reach it through continuous seed mutation. Angora uses taint tracking on the dynamic data flow to find the vulnerability in the program. CAFL uses a set of ordered crash-triggering target locations as targets for DGF. As CAFL is not open source, only its crash reproduction time on the LAVA-1 dataset is compared with that of the CONFF fuzzer.

\textbf{Target crashes.}
We evaluated the crash replication capacity of the CONFF fuzzer on the widely adopted LAVA-1 dataset. We also selected some vulnerabilities with crash reproduction information from CVEs to evaluate the applicability of the CONFF fuzzer. During fuzzing, the CONFF fuzzer additionally discovered some crashes mentioned in GitHub issues and some unknown crashes, and we also made performance comparisons on these crashes.

\textbf{Initial seeds.} 
The CONFF fuzzer relies on byte-level dynamic data flow tracking. Some path constraints can only be tracked with seeds meeting the format requirements. For random inputs, satisfying untrackable implicit paths can be time consuming. Therefore, the CONFF fuzzer uses some simple formatted input files as initial seeds to fuzz the tested program. The same initial seeds are used for the baseline fuzzers.

\textbf{Crash location.}
Directed greybox fuzzers guide seed mutation with crash location information. As shown in figure \ref{patchloc}, crash location information can be extracted through three ways, including AddressSantizer, Manual, and Git diff. During crash reproduction, any type of crash location information can be input into the fuzzer to determine the target location.

\textbf{Experimental environment.} The CONFF fuzzer runs each tested program in the same system environment. Specifically, we extracted the LLVM IR bitcode of the programs using LLVM 12.0.1 on Ubuntu 18.04 docker image. The program was compiled using clang 12.0.1 and run on Ubuntu 20.04 with an Intel i7-10700 CPU @ 2.90GHz and 16GB RAM.

\subsection{Crash reproduction on the LAVA-1 dataset}\label{LAVA1}
LAVA-1 is a dataset with bugs manually injected into the program files, only one bug per program. As LAVA-1 provides detailed crash information, we obtain the provided CRASH\_INPUT crash file and collect the stack track of the crash as control flow information to guide the mutation of the CONFF fuzzer. Fuzzing results are presented in Table \ref{LAVA-1}. We can see that for most crashes in the LAVA-1 dataset, the CONFF fuzzer triggers crashes faster than state-of-the-art fuzzers (AFLGo, Angora, and CAFL). \par

The results also show that the crash reproductions are generally fast on the LAVA-1 dataset as the distance between the seed in ELF format and the crash location (i.e., the target location) is short in LAVA-1, and the directed fuzzers can reach the target location quickly. We also experimented with randomly generated seeds as input, and the results showed that the crash reproduction speed was greatly reduced. Analysis reveals the reason as the implicit control flows difficult for random seeds to pass through. Formatted seeds can meet the data conditions of these implicit control flows, thus speeding up the crash reproduction. This also demonstrates the advantage of using formatted file as the initial seeds. \par

\subsection{Reproduction of real-world crashes}\label{real}
We compared the crash reproduction speed of 18 vulnerabilities involved in CVEs, as shown in Table \ref{CVE}. We set the timeout to 24 hours. All target location information is generated with the given sanitizer information. We determine whether the vulnerability is successfully reproduced based on the path and crash type reported by AddressSanitizer. Note that since the stack information of \textit{jasper} cannot be collected, whether the program is aborted servers as an indicator of crash reproduction. We performed multiple rounds of fuzzing and used the shortest time to reproduce the crash for evaluation. \par

The fuzzing results are presented in Table \ref{CVE}. Compared with AFLGo and Angora, the CONFF fuzzer achieved the highest speed in 13 out of 18 crash reproductions. Overall, the CONFF fuzzer outperforms AFLGo by 27.3 times and Angora by 23.7 times on average. The experimental results confirm that the CONFF fuzzer combining the control flow and data flow information for path constraint filtering and focusing can effectively improve the efficiency of DGF. \par

As shown in Table \ref{newCrash}, we also discovered 15 additional crashes during the fuzzing on the real-world dataset. Some crashes are mentioned on github issues, while "-" represents undetected 0-day crashes. 
Reproduction of these crashes are also attempted with AFLGo and Angora.
Similarly, we use the stack track of crashes provided by github issues or AddressSanitizer as crash location information for the fuzzers. The experimental results are presented in Table \ref{newCrash}.
In terms of the new crashes, the CONFF fuzzer shows more advantages over AFLGo and Angora, and AFLGo and Angora fail to reproduce many crashes even with timeouts. \par

\subsection{Performance Evaluations}\label{related}

\begin{figure}[htpb]
	\centering
	\includegraphics[width=0.5\textwidth]{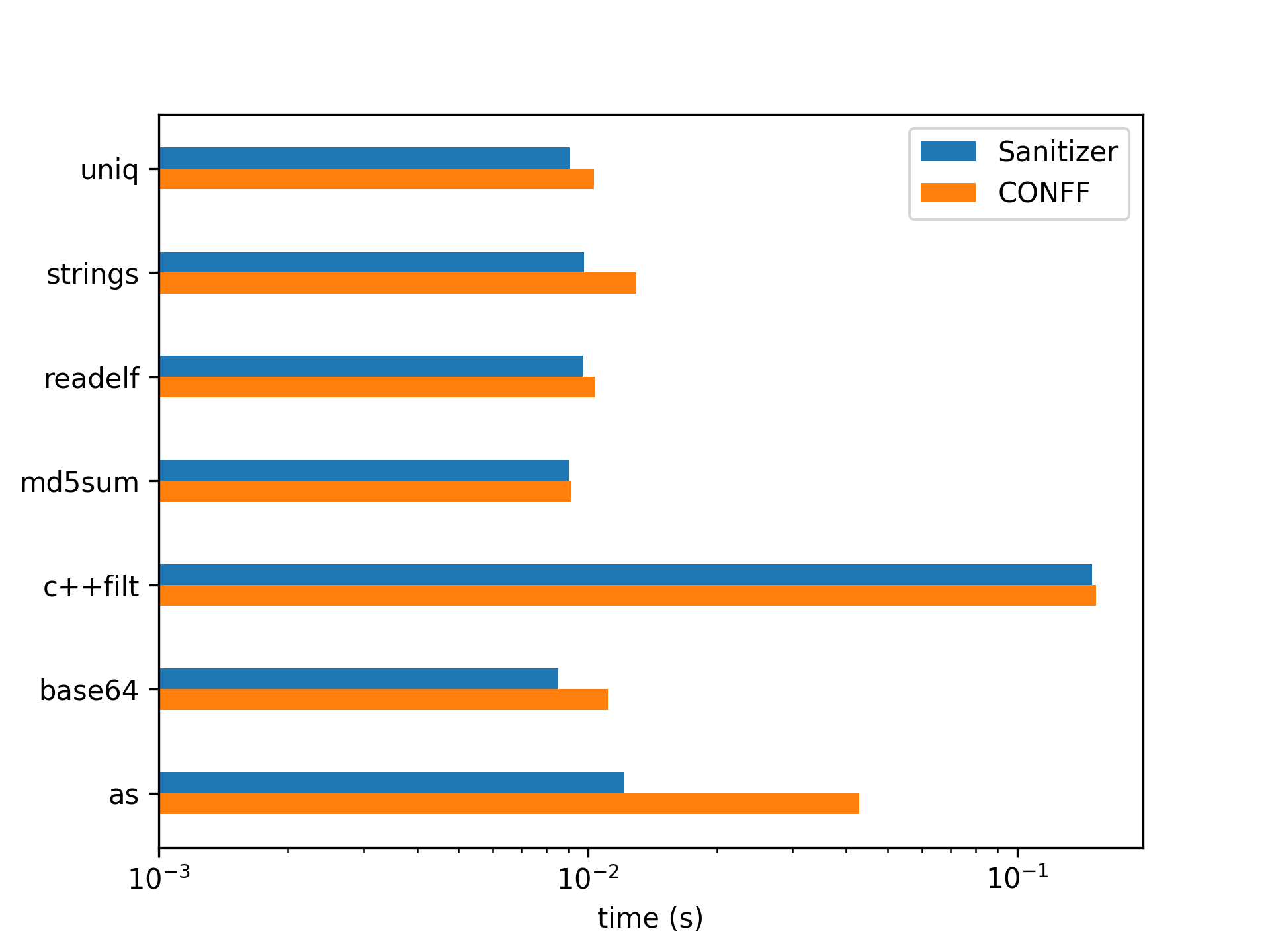}  
	\setlength{\abovecaptionskip}{-10pt}
    \setlength{\belowcaptionskip}{-5pt}
	\caption{Time cost comparison.}
	\label{performance}
\end{figure}

\textbf{Time cost comparison.}
As the CONFF fuzzer is implemented using the Clang Sanitizer, it serves as a benchmark to evaluate the efficiency of the CONFF fuzzer. 
We selected 7 open source projects in different fields and compiled them with clang sanitizer and the CONFF fuzzer, respectively. For each pair of generated programs, the same seed with a length of 1 kB was fed, and the execution time was recorded. Each tested program was executed 10 times to obtain the average value. 
As shown in Figure \ref{performance}, DGF and the sanitizer spend approximately the same execution time for most projects. On the \textit{as} program, the CONFF fuzzer has 3.75 times the time cost.
On average, the time cost of using the CONFF fuzzer to obtain dynamic data flows is 1.53 times that of Clang Sanitizer.

\begin{figure}[htpb]
	\centering
	\includegraphics[width=0.5\textwidth]{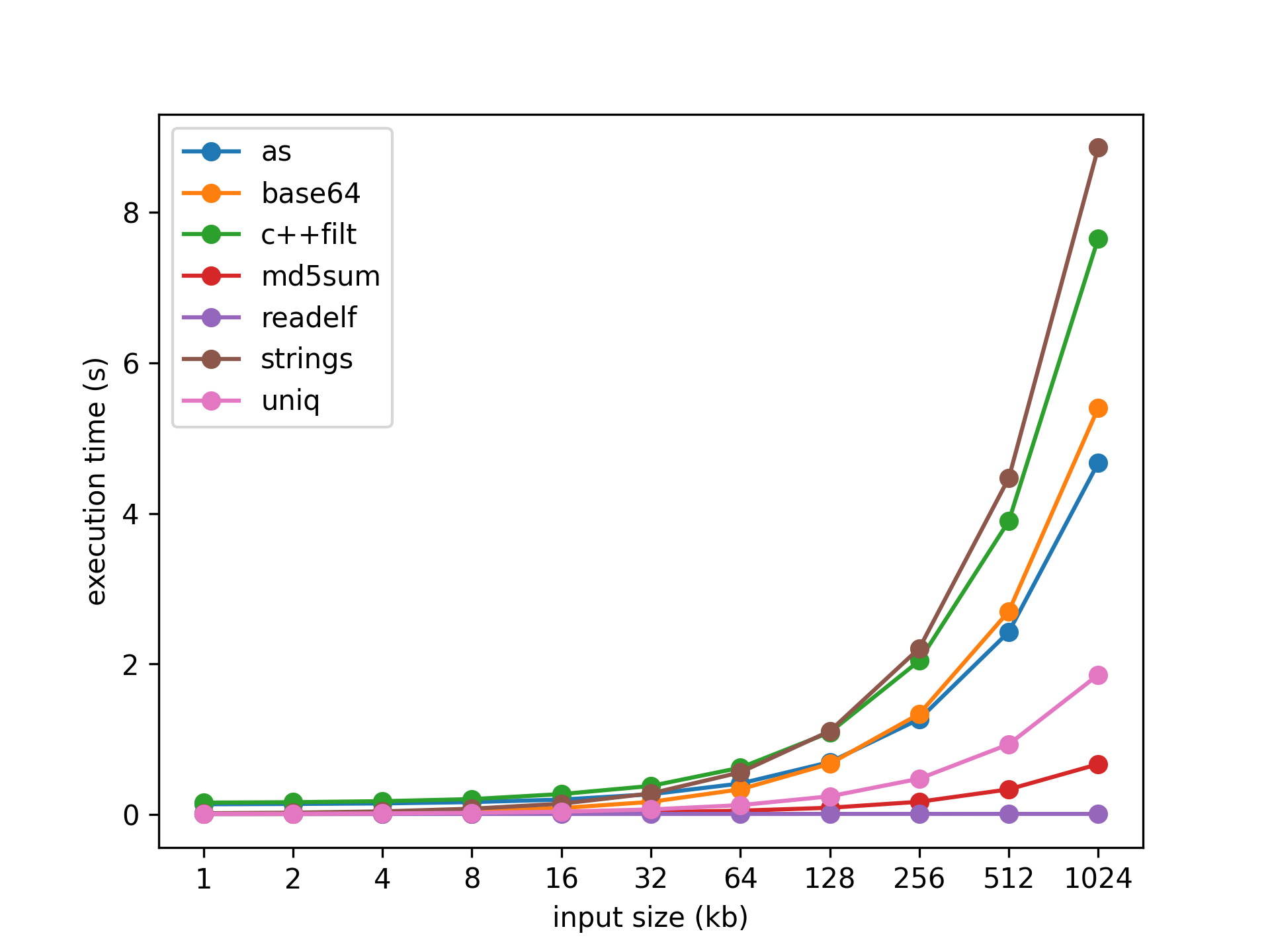}
	\setlength{\abovecaptionskip}{-10pt}
    \setlength{\belowcaptionskip}{-5pt}
	\caption{Execution time for different input lengths.}
	\label{seedlength}
\end{figure}

\textbf{Time cost for different input lengths.} 
Figure \ref{seedlength} shows the relationship between the seed length and the execution time of the tested program compiled with CONFF. In the case that the input seed can be parsed by the program, a greater seed length increases the number of code blocks and path constraints passed by the program while prolonging the execution time. If the input seed cannot be parsed, the program goes through the exception handling branch, and the execution time will not increase, as shown by \textit{readelf}. Therefore, it is necessary to limit the seed length when using dynamic data flows for fuzzing, and it is generally recommended that the initial seed size be less than 4 MB.

\begin{figure}[htpb]
	\centering
	\includegraphics[width=0.55\textwidth,height=0.46\textheight]{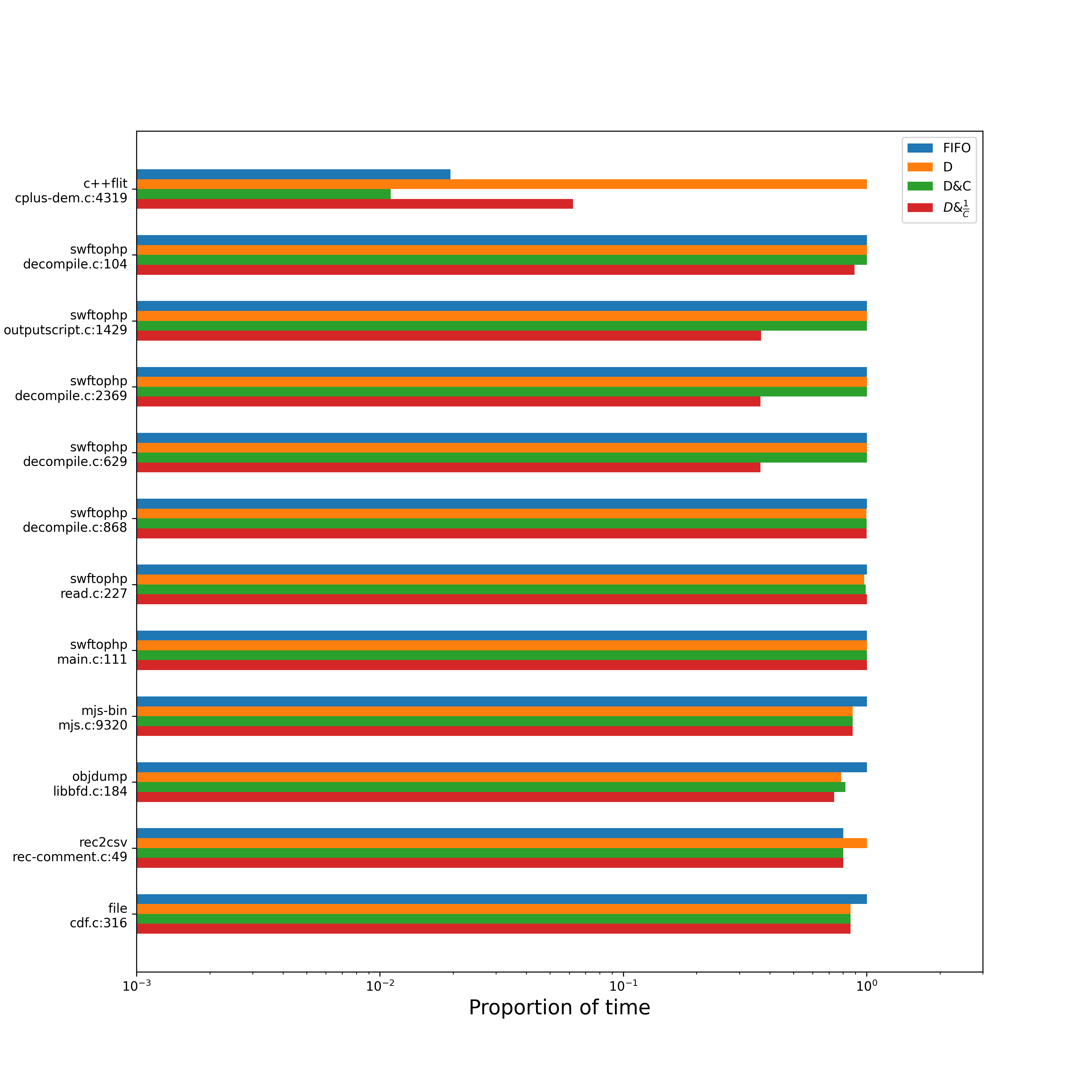}
	\setlength{\abovecaptionskip}{-15pt}
    \setlength{\belowcaptionskip}{-10pt}
	\caption{Seed priority on triggering crashes.}
	\label{priority}
\end{figure}

\textbf{Seed priority for triggering crashes.} 
To demonstrate the effect of seed-priority scheduling on reaching the crash location, we compared different scheduling strategies for the seed. We selected 12 projects and adopted 4 seed priority indicators for fuzzing, including arrival time (FIFO), Distance ($D$), Distance\&Coverage($D\&C$), and Distance\&1/Coverage($D\&\tfrac{1}{C}$). Each priority indicator in each project is tested 3 times, and the average time elapsed for triggering the crash is calculated. To facilitate display and comparison, we take the maximum time elapsed in each project and divide each time elapsed by the maximum value. Figure \ref{priority} shows the scaled time elapsed, where the $D\&\tfrac{1}{C}$ priority scheme exhibits certain advantages in most projects. This experimental result also confirms our previous hypothesis that a higher path coverage means a higher possibility of triggering a crash. Admittedly, the seed distance is still the main factor in the $D\&\tfrac{1}{C}$ priority scheme, and the coverage rate is secondary. Thus, the experimental results suggest that the coverage in some projects has no chance to affect seed scheduling.

\section{Related Work}

\textbf{Types of directed fuzzers.}
Classical directed greybox fuzzers strive to reach the target location of the tested program through seed mutation.
AFLGo\cite{bohme2017directed} and Hawkeye\cite{chen2018hawkeye} were proposed to direct the seed to a manually defined target program location consisting of a file name and a line number.
ParmeSan\cite{osterlund2020parmesan} directs the seed mutation to drive the program to the location of the Sanitizer, thus more effectively exposing possible vulnerabilities in the program.
CAFL\cite{lee2021constraint} automatically generates manually defined templates to satisfy the data conditions of the sorted target sites.
Other directed fuzzers target specific metrics.
SlowFuzz\cite{petsios2017slowfuzz} and PERFFuzz\cite{lemieux2018perffuzz} use resource utilization as a target to find algorithmic complexity vulnerabilities. UAFL\cite{wang2020typestate} and UAFuzz\cite{nguyen2020binary} target the use-after-free vulnerabilities. 
TIFF\cite{jain2018tiff} and SAVIOR\cite{chen2020savior} mainly target vulnerabilities triggering memory corruptions.
RGF\cite{zhu2021regression} focuses on the fuzzing of codes with more recent or more frequent changes.

\textbf{Target identification.}
Selecting the target location is an essential task for directed fuzzing, and the target location is usually marked manually.
Researchers have adopted various methods to collect information on the desired target location.
1) Obtaining the line numbers of changed codes from the source code commit log. RGF\cite{zhu2021regression} acquires the line numbers of recently or frequently changed codes and directs the seed to the location of frequent code changes. 2) Obtaining the relevant information from the patch related locations. KATCH\cite{marinescu2013katch} and CAFL\cite{lee2021constraint} can extract target locations from the patch information for further fuzzing. 3) Using the generated stack track as the location information. AFLGo\cite{bohme2017directed} uses the Sanitizer-generated stack track as the target location. 

\textbf{Distance measurements.}
It is generally believed that seeds closer to the target location have higher probabilities of reaching it.
AFLGo\cite{bohme2017directed} uses Dijkstra's algorithm to calculate the shortest path to the target location and the harmonic mean to define the difference between node distances in the case of multiple targets. 
Hawkeye\cite{chen2018hawkeye} and Berry\cite{liang2020sequence} use the similarity between the seed execution trace and the target execution trace to measure the distance between the seed and the target location.
An ordered path distance measurement has been proposed in CAFL\cite{lee2021constraint}. Taking the use-after-free vulnerability as an example, the free function is usually not on the shortest path to the target location. Therefore, the vulnerability path must first trigger the \textit{free} function and then reach the target location of \textit{use} function to correctly reproduce the vulnerability.


\textbf{Seed evaluation metrics.}
AFL\cite{AFL} uses code coverage, one of the most widely used and readily available metrics, as the input seed evaluation metric.
Most directed fuzzers, such as AFLGo\cite{bohme2017directed}, use the distance to the target location as the seed evaluation metric. 
V-Fuzz\cite{li2019v} utilizes function vulnerability probabilities generated by graph embedding networks to predict the likelihood of an input seed triggering a vulnerability.

\textbf{Seed selection.} In terms of seed selection, AFLGo\cite{bohme2017directed} uses a simulated annealing-based power schedule algorithm to gradually allocate more time to seeds closer to the target location for fuzzing. Hawkeye\cite{chen2018hawkeye} adds a prioritization method to simulated annealing. 
Greyone\cite{gan2020greyone} uses three seed queues for seed priority scheduling.
CAFL\cite{lee2021constraint} attaches different probabilities to the seeds and uses their probabilities for selection.
RETECS\cite{spieker2017reinforcement}, RLCheck\cite{reddy2020quickly}, and AFL-HIER\cite{wang2021reinforcement} use reinforcement learning to prioritize and schedule the seeds. 
SmartSeed\cite{lyu2018smartseed} uses machine learning methods to generate valid input seed sets.
DeepHunter\cite{xie2019deephunter} and Learn\&Fuzz\cite{godefroid2017learn} use deep learning to generate the format-specific input seed set.

\section{Conclusion}
In this paper, we proposed a directed greybox fuzzer with CONFF to reach the target location of the tested program faster.
Through constraint filtering and selection, the CONFF fuzzer only focused on the path constraint most favorable to quickly reaching the target location. By mapping the data conditions of the path constraint to the seed bytes, the CONFF fuzzer precisely focused on the associated seed bytes and quickly satisfied the path constraints through strategic mutations. Combining constraint focusing and seed bytes focusing, the CONFF fuzzer significantly improved the efficiency of reaching the target location. Moreover, through the scheduling of seeds and constraints, the CONFF fuzzer covered as many paths as possible to reproduce the target crash.\par

We implemented the CONFF fuzzer and evaluated it on the LAVA-1 dataset and some real-world crashes. Experimental results confirmed the feasibility of the CONFF fuzzer in reproducing crashes, which outperformed the state-of-the-art fuzzers. \par

\bibliographystyle{plain}
\bibliography{reference}

	
\end{document}